\documentclass[]{spie}

\usepackage[]{graphicx}
\usepackage{subfigure}
\usepackage{color}
\usepackage{fig}

\title{Gemini Planet Imager Observational Calibrations IV: Wavelength Calibration and 
Flexure Correction for the Integral Field Spectrograph} 

\author{Schuyler G. Wolff\supit{a},  Marshall D. Perrin\supit{b}, Jerome Maire\supit{c}, Patrick J. Ingraham\supit{d}, Fredrik T. Rantakyr\"o\supit{e}, and Pascale Hibon\supit{e}, with the GPI team.
\skiplinehalf
\supit{a}Johns Hopkins University, 3400 North Charles St., Baltimore, MD, USA; \\
\supit{b}Space Telescope Science Institute, 3700 San Martin Drive, Baltimore, MD, USA;  \\
\supit{c}Dunlap Institute for Astronomy \& Astrophysics, 50 George St., Toronto, ON, Canada; \\
\supit{d}Stanford University, 295 Galvez St., Stanford, CA, USA; \\
\supit{e}Gemini South Observatory, Casilla 603, La Serena, Chile	
}

\authorinfo{Further author information: (Send correspondence to S.G.W.)\\  S.G.W.: E-mail: swolff@stsci.edu, Telephone: 1 410 338 2428 }

\begin{document} 
  \maketitle 

\begin{abstract}
We present the wavelength calibration for the lenslet-based Integral Field Spectrograph (IFS) that serves as the  science instrument for the Gemini Planet Imager (GPI). The GPI IFS features a 2.7'' x 2.7'' field of view and a 190 x 190 lenslet array (14.3 mas/lenslet) operating in \textit{Y, J, H}, and \textit{K} bands with spectral resolving power ranging from $R$ $\sim$ 35 to 78. Due to variations across the field of view, a unique wavelength solution is determined for each lenslet characterized by a two-dimensional position, the spectral dispersion, and the rotation of the spectrum with respect to the detector axes. The four free parameters are fit using a constrained Levenberg-Marquardt least-squares minimization algorithm, which compares an individual lenslet's arc lamp spectrum to a simulated arc lamp spectrum. This method enables measurement of spectral positions to better than 1/10th of a pixel on the GPI IFS detector using Gemini's facility calibration lamp unit GCAL, improving spectral extraction accuracy compared to earlier approaches. Using such wavelength calibrations we have measured how internal flexure of the spectrograph with changing zenith angle shifts spectra on the detector.  We describe the methods used to compensate for these shifts when assembling datacubes from on-sky observations using GPI.  
\end{abstract}


\keywords{Gemini Planet Imager, GPI, Integral Field Spectrograph, Wavelength Calibration}

\section{INTRODUCTION}
\label{sec:intro}  

The science instrument for the Gemini Planet Imager is the Integral Field Spectrograph (IFS) operating in the near-IR.\cite{larkin} The IFS uses a lenslet-based design and a HAWAII-2RG detector. The instrument has a $\sim$ 2.7'' x 2.7'' field of view partitioned by a $\sim$ 190 x 190 lenslet array. The GPI IFS contains five bandpasses (\textit{Y, J, H, K1} and \textit{K2}) that has a spectral resolving power of $R$ $\sim$ 35 - 78 depending on the band. \textit{K} band was split to allow all 36000 lenslet spectra to fit on the detector.\cite{2014AAS22320204C} The relatively low spectral resolution of GPI allows for the small lenslet plate scale of 14.3 mas/lenslet necessary for Nyquist sampling at the shortest wavelengths while providing enough detail to distinguish between planetary atmospheric models.\cite{2011PASP123692M} 
A data reduction pipeline has been developed by the GPI team to process this complex array of micro-spectra and has been made publicly available.\cite{2014AAS22334814P,perrinpipe} 

The focus of this paper, and one of the main obstacles in calibrating the GPI IFS, is the wavelength calibration. Each reimaged lenslet has a unique position on the detector which changes with time and elevation resulting from gravitationally induced shifts due to flexure within the IFS, and distinct spectral properties. Calibrations for all Gemini South instruments are performed using the Gemini Facility Calibration Unit (GCAL)\cite{1997SPIE.2871.1171R} that occupies one of the ports on the bottom of the Gemini Telescope. A fold mirror directs light from GCAL into GPI. The Gemini Planet Imager has no internal wavelength calibration source and must rely on GCAL for all wavelength calibrations. GCAL contains four arc line lamps, but only Ar and Xe are useful for spectral calibration for GPI; the spectral lines for the CuAr and ThAr lamps are too faint. While Xe has the advantage of fewer blended lines, the Xe lamp is 3 - 20 times fainter and requires more overhead time for calibrations. The performance of Xe and Ar lamps are discussed in Section \ref{sec:arvsxe}. 

GPI was installed at Gemini South in October 2013, and has now completed five observing runs as of May 2014 including a successful early science run. Throughout the runs, GPI has performed well and has already produced some interesting scientific results. \cite{2014AAS22322902M, 2014arXiv1403.7520M} 
Here we present a wavelength calibration algorithm written as a module within the GPI Data Reduction Pipeline and tested using first light results of GPI. We aim to produce an accurate wavelength solution for each lenslet across the field of view for science data with the minimal amount of overhead calibration time. 
A description of the wavelength solution algorithm used for GPI is given in Section \ref{sec:wavecal}. Both centroid and least squared algorithms are presented. The performance and accuracy of the wavelength solution is discussed in Section \ref{sec:results}. The observed flexure within the Integral Field Spectrograph is addressed in Section \ref{sec:flexure}. 

\section{WAVELENGTH CALIBRATION}
\label{sec:wavecal} 

The Integral Field Spectrograph for GPI produces $\sim 36000$ spectra each with a unique position and spectral properties that differ measurably across the field of view. Spectra are tightly packed on the detector plane, with a separation of approximately 4.5 pixels in the cross-dispersion direction. Furthermore, at the low resolving power of GPI the lines from the Xe and Ar lamps often appear strongly blended with few isolated peaks. All of these factors contribute to the need for a flexible and reliable wavelength calibration algorithm. 

We begin with a dark subtracted and bad pixel corrected lamp image, and examine each lenslet spectrum individually.
The wavelength as a function of position for a given lenslet is represented as a line and defined by Equation \ref{eq:lambda}. 


\begin{equation}
x = x_{0} + \sin{\theta} \frac{\lambda - \lambda_{0}}{w} \,\,\,\,\,\, \mathrm{and} \,\,\,\,\,\, y = y_{0} - \cos{\theta} \frac{\lambda - \lambda_{0}}{w}
\label{eq:lambda}
\end{equation}

\noindent Here $\lambda$ is the wavelength in microns, $w$ is the dispersion in $\mu$m/pixel,  $x$ and $y$ are pixel positions on the detector, $\lambda_{0}$ is some reference wavelength (in $\mu$m), and $x_{0}$ and $y_{0}$ gives the pixel locations for $\lambda_{0}$. These values are calculated individually for each lenslet and saved in a 281 x 281 x 5 datacube. These data cubes are later used to extract science spectral data into 37 wavelength channels. In the sections below, we describe the methods used to calculate these values for a given lenslet spectrum.

\subsection{Centroiding Algorithm}

The original algorithm developed for wavelength calibration for GPI worked by measuring the positions of individual spectral lines one at a time and then fitting Equation 1 to the derived positions of each line. The routine begins with a 2D detector arc lamp image, locates the brightest spectral peak in the central lenslet. For that lenslet, it measures the location of the spectral peaks for a predefined set of emission lines using a barycenter algorithm. After fitting the central lenslet, the code works its way outwards fitting each lenslet across the detector. The position of each subsequent lenslet is estimated by calculating an offset from the prior fit lenslet based on assumed values for the separation and orientation of the lenslets. Once the positions of the individual spectra were calculated, the dispersion and tilts for each lenslet are reevaluated. 

\begin{figure}
    \begin{center}
 \includegraphics[width=0.3\textwidth]{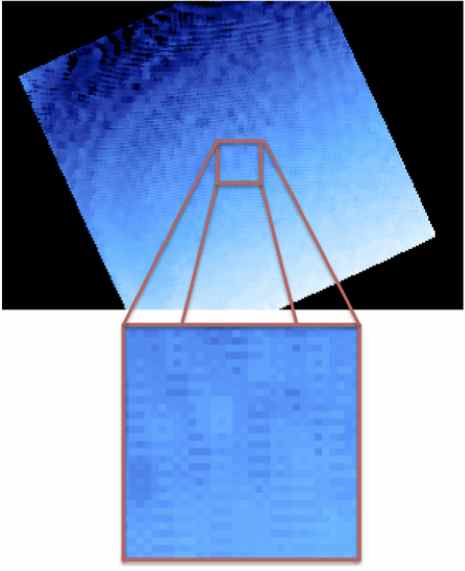}
     \end{center}
    \caption[example] 
   { \label{fig:oldmethod} 
   Illustration of the moir\'{e} pattern. The top image gives the dispersion of all 36000 spectra. This should be a smooth distribution, but the zoomed in region shows the moir\'{e} pattern caused by aliasing between adjacent lenslets. }
  \end{figure}

Using this method, we found that 99.9 \% of spectra are detected with an accuracy better than 0.3 pixels
and 80 \% are detected within 0.12 pixels in lab testing of arc lamp data.\cite{testgpipaper} However, we discovered that the asymmetric shape of the lenslet PSFs coupled with the spectral peaks being under Nyquist sampled led to errors in the spectral positions found using the center-of-mass centroiding algorithm. These errors result in different wavelength solutions between adjacent spectra causing the moir\'{e} pattern seen in Figure \ref{fig:oldmethod}.

\subsection{Least Squares Fitting Algorithm}
\label{sec:leastsqrs}

   \begin{figure}[t]
    \begin{center}
   \subfigure[H band Xe Arc Lamp]
	{
     \includegraphics[width=0.4\textwidth]{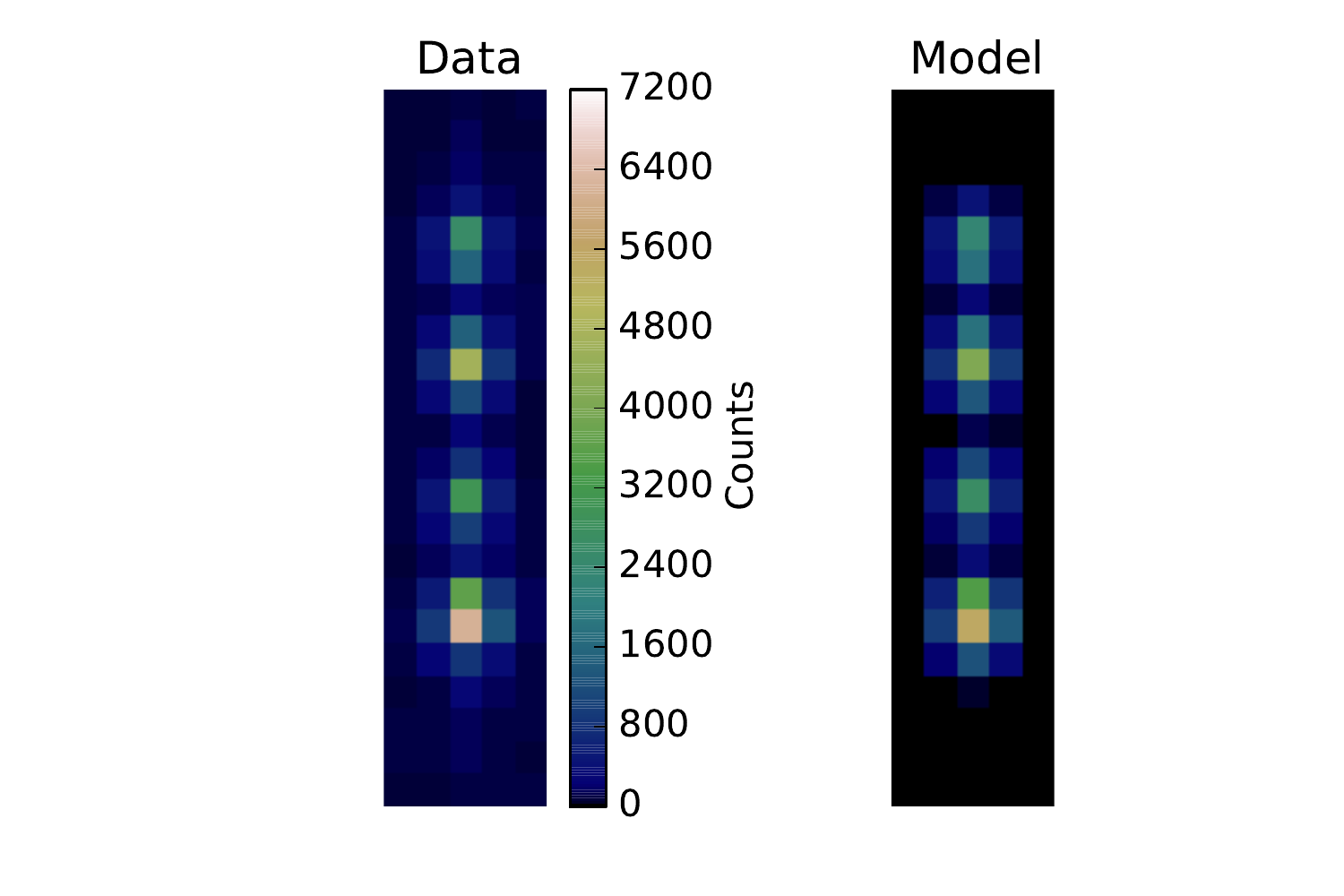}
     \includegraphics[width=0.35\textwidth]{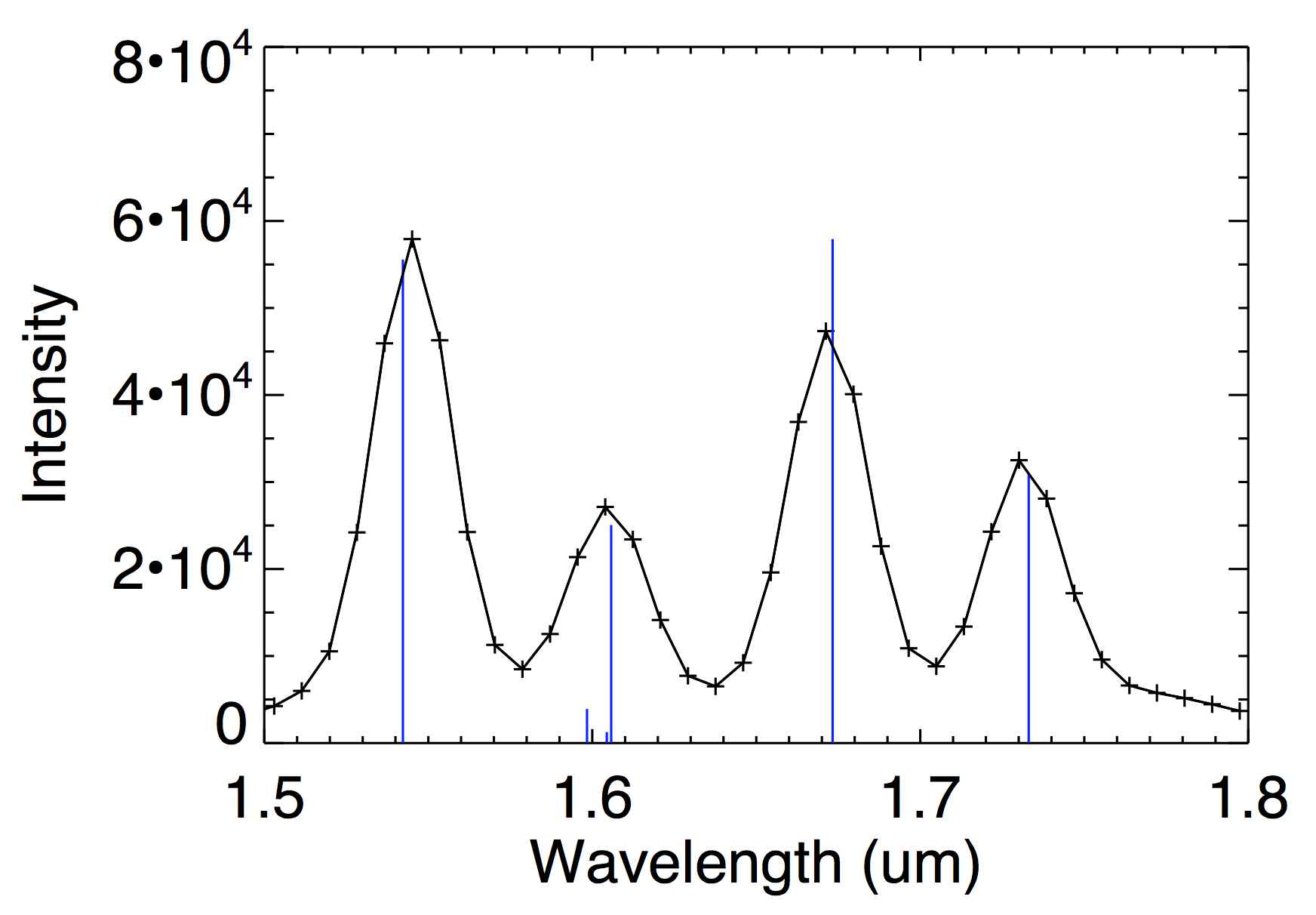}
     \label{fig:xedata}
	}  
   \subfigure[H band Ar Arc Lamp]
   {
   \includegraphics[width=0.4\textwidth]{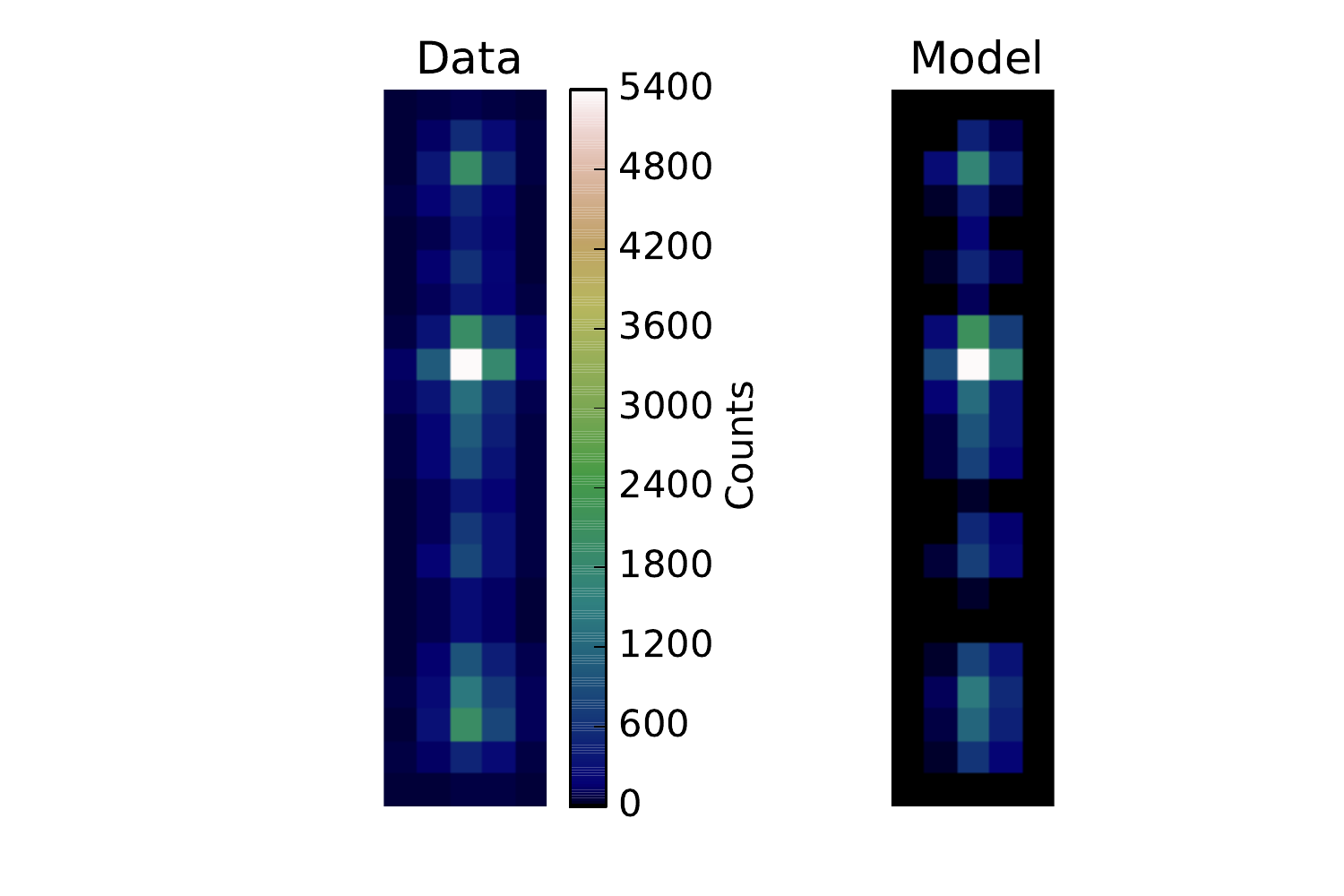}
        \includegraphics[width=0.35\textwidth]{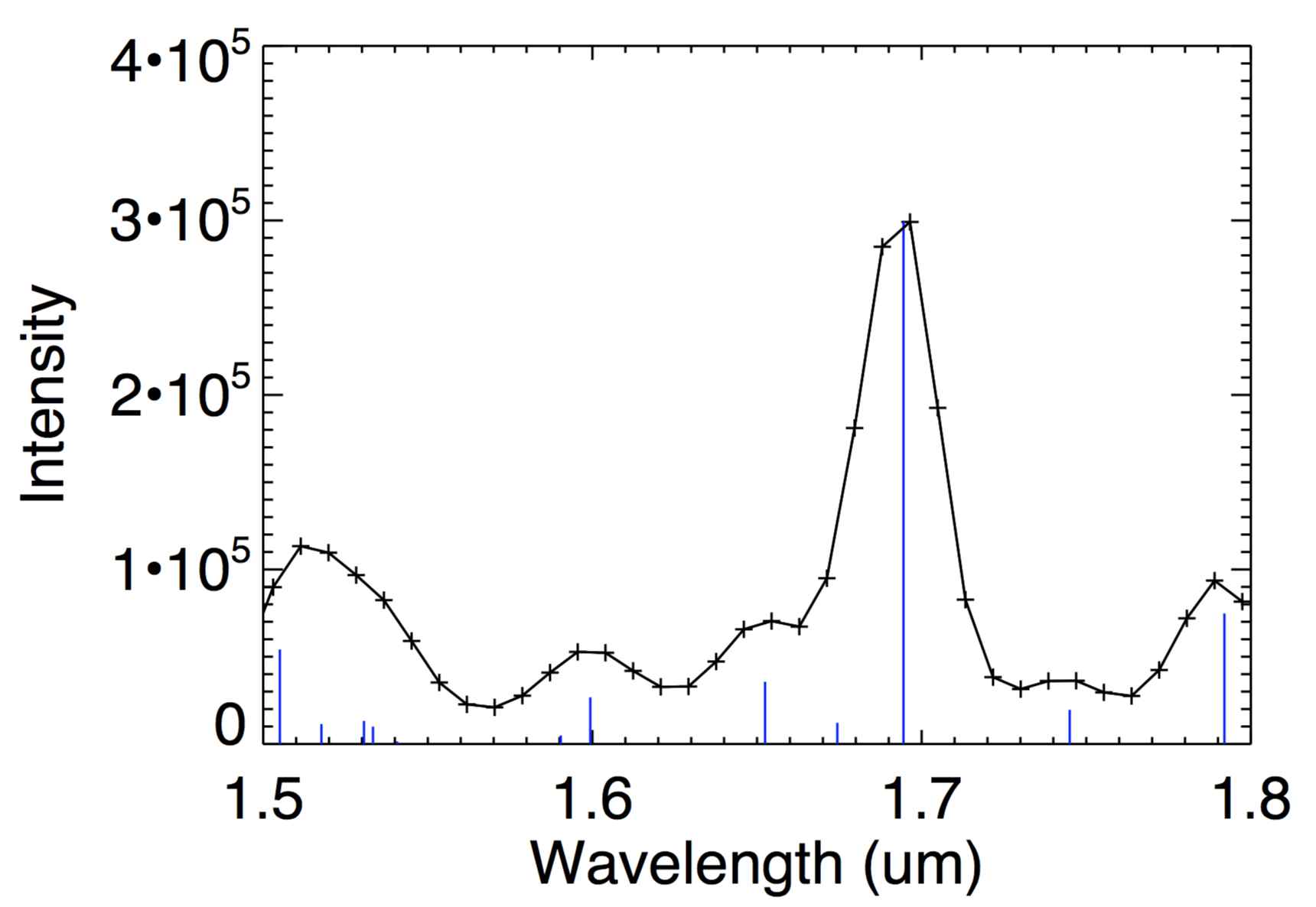}
   \label{fig:ardata}
   }
   \end{center}
   \caption[example] 
   { \label{fig:images} 
    Illustration of the least squares fitting results. For each lenslet, the 2D observed spectrum and compared to a modeled lenslet spectrum. (a) H band + Xe arc lamp; On the left is an observed lenslet spectrum with best fit modeled spectrum plotted on the same scale. The spectrum used to model the Xe arc is given on the right. The GCAL emission lines are plotted in blue while the black line and points give the Xe spectrum binned to the resolution of the GPI IFS. (b) H band + Ar arc lamp; same as (b). The Ar lamp is much more difficult to fit because there are many blended emission lines and only a single sharp peak. The $J$ band Ar arc lamp is even more difficult. }
  
  \end{figure} 

In order to correct the issues with the centroiding algorithm, a new method was implemented designed to fit all the peaks in the lenslet spectrum simultaneously, and with increased sub-pixel sensitivity. The new algorithm uses a least squares fitting approach to compare an individual lenslet spectrum in the 2D detector plane to a modeled spectrum. We implement this using the mpfit2dfun IDL package written by Craig B. Markwardt, which fits parameters $P$ for a user defined function $f ( x_{i}, P )$ using the Levenberg-Marquardt algorithm. The Levenberg-Marquardt method is an non-linear least squares fitting technique which aims to minimize the error weighted squared residuals.\cite{Levenberg,marquardt} 

\begin{equation}
\mathrm{min}_{P} \displaystyle\sum\limits_{i=0}^{M} r_{i}(P)^2, \,\,\,\,\,\,\,\,\, \mathrm{where} \,\,\, r_{i}(P) = \frac{y_{i} - f ( x_{i}, P )}{\sigma_{i}}
\end{equation}

In this case, the user defined function returns an array containing $N$ Gaussian PSFs of varying peak flux and FWHM, where the number of peaks in a lenslet spectrum, $N$, varies with band and filter combination. For example, an \textit{H} band Xe lamp exposure fits four Gaussians simultaneously at the four peak locations resolved in a lenslet spectrum shown in Figure \ref{fig:xedata}. The more complicated case of an \textit{H} band Ar lamp exposure is illustrated in Figure \ref{fig:ardata} with only one clear spectral peak and many blended lines. For this case, there are twelve emission lines in this band, which we approximating by fitting only the six brightest emission lines.  The free parameters in the fit are the initial $x_{0}$ and $y_{0}$ values, the dispersion ($w$), the angle ($\theta$) of rotation for the lenslet, the flux ratios of the peaks, and the total flux scaling for the lenslet. A previous wavelength solution is read in and used as an initial guess for the starting parameters of the fit. Constraints on acceptable values can be placed on each of the free parameters. The errors ($\sigma_{i}$) are given by the photon noise and weighted by the bad pixel map. 
	
Empirically this algorithm does succeed in mitigating the problems  which impacted the centroiding algorithm. Known bad pixels can be weighted to zero to avoid affecting the fits, and errors in fitting one lenslet do not propagate into erroneous starting guesses for subsequent lenslet fits. Statistical tests of the derived lenslet locations show reduced statistical biases compared to the centroiding algorithm; specifically the histogram for position offsets is much more Gaussian and the bimodal distribution (Moir\'{e} pattern) is greatly reduced. The wavelength solution can be calculated by using the ``2D Wavelength Solution" module in the GPI Data Reduction Pipeline. 	
	
	\newpage
		
\subsection{Gaussian vs. Microlens PSFs}

\begin{figure}
     \centering
      \subfigure[H Band Xe Data]
	{
     \includegraphics[width=0.3\textwidth]{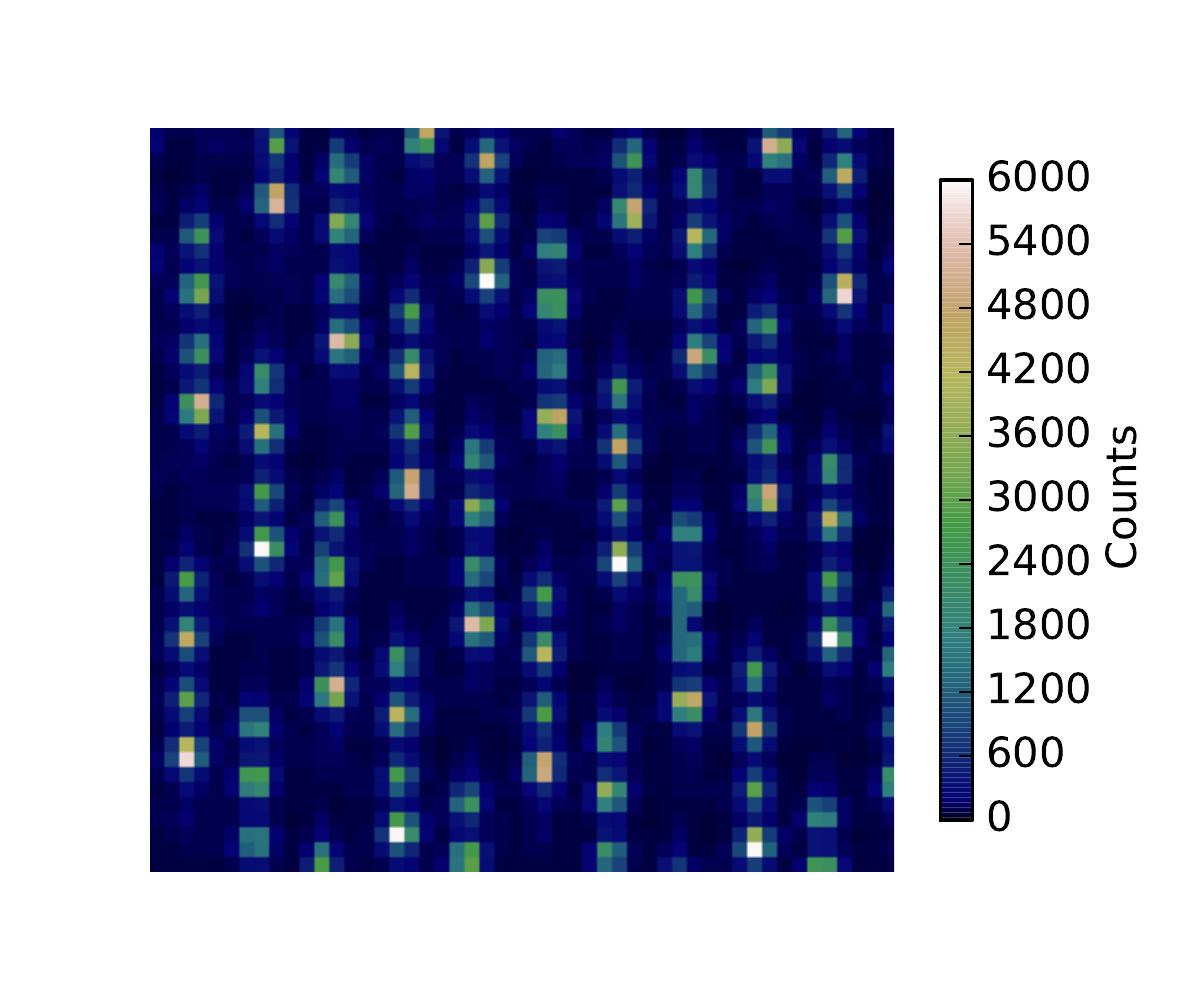}
     \label{fig:gaussres}
   }
   \subfigure[Gauss PSF Model]
   {
   \includegraphics[width=0.25\textwidth]{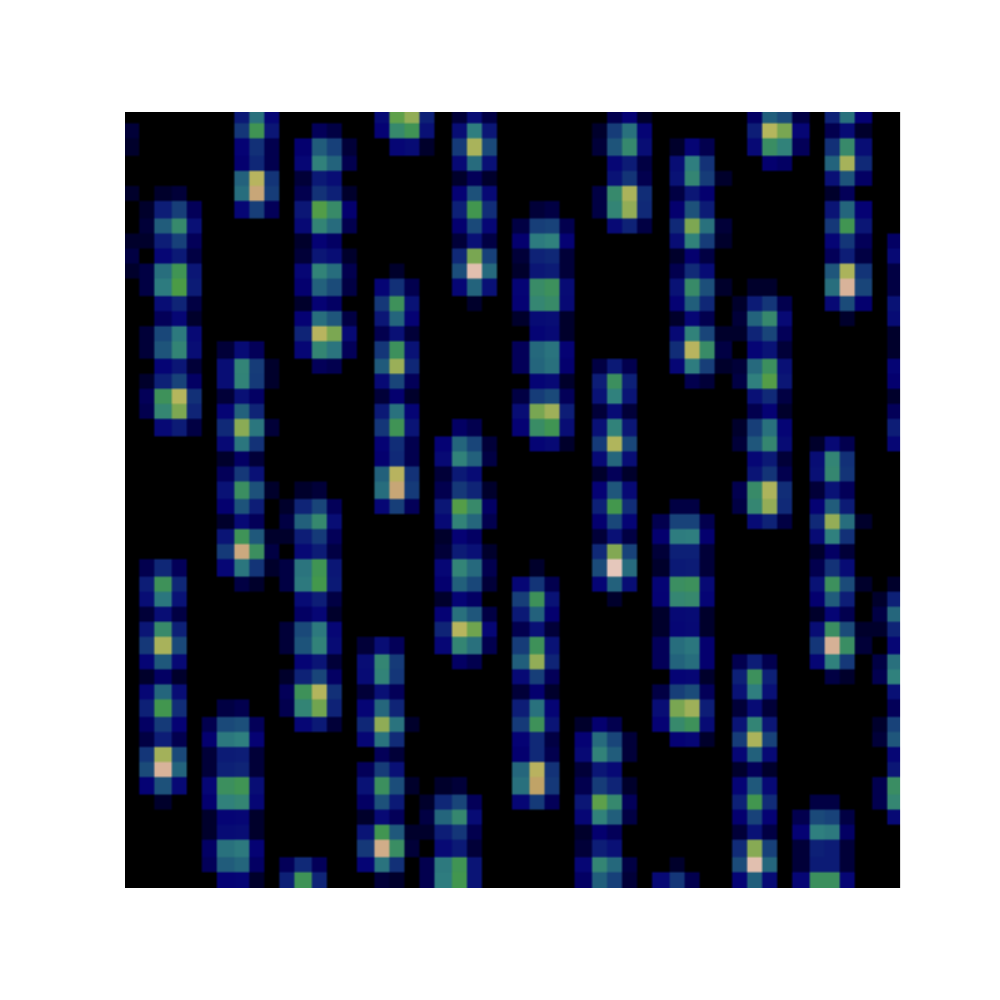}
   \label{fig:microlensres}
   }
   \subfigure[Microlens PSF Model]
    {
   \includegraphics[width=0.25\textwidth]{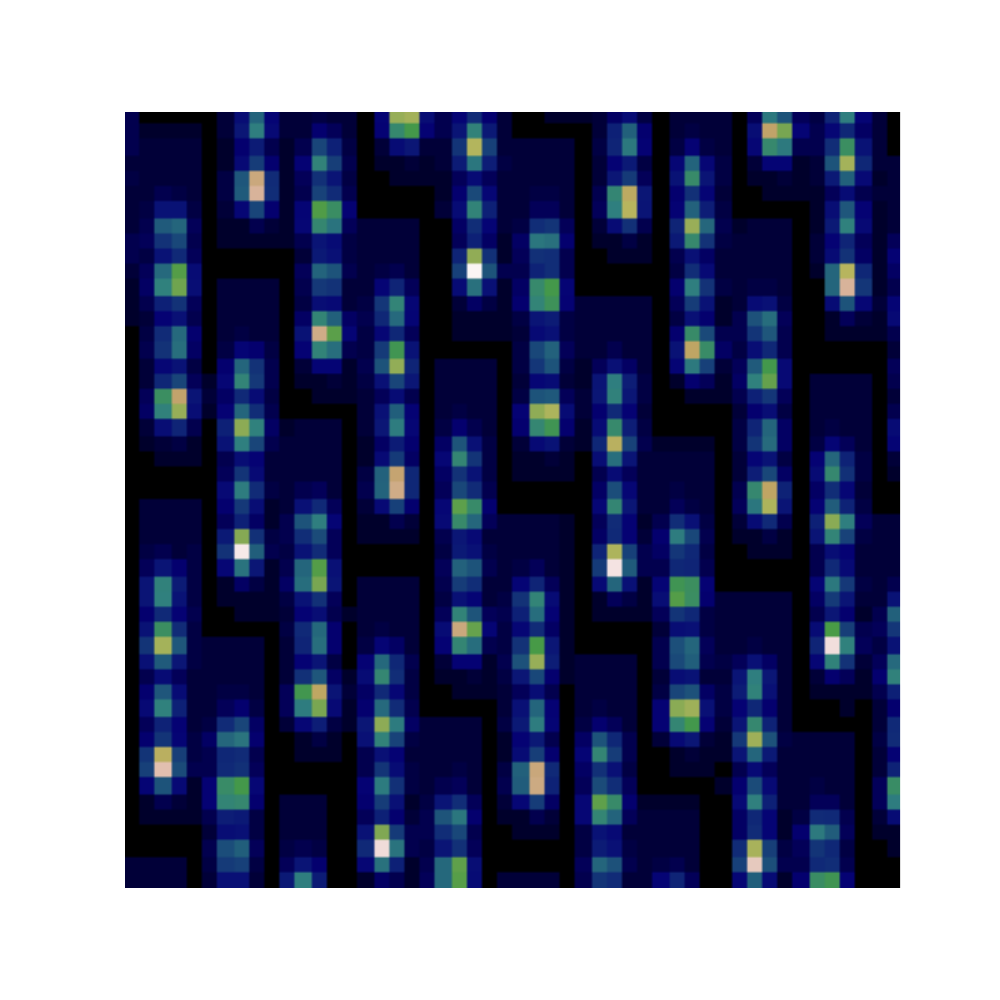}
   \label{fig:microlensres}
   }
      \subfigure[Gauss PSF Residuals]
   {
   \includegraphics[width=0.3\textwidth]{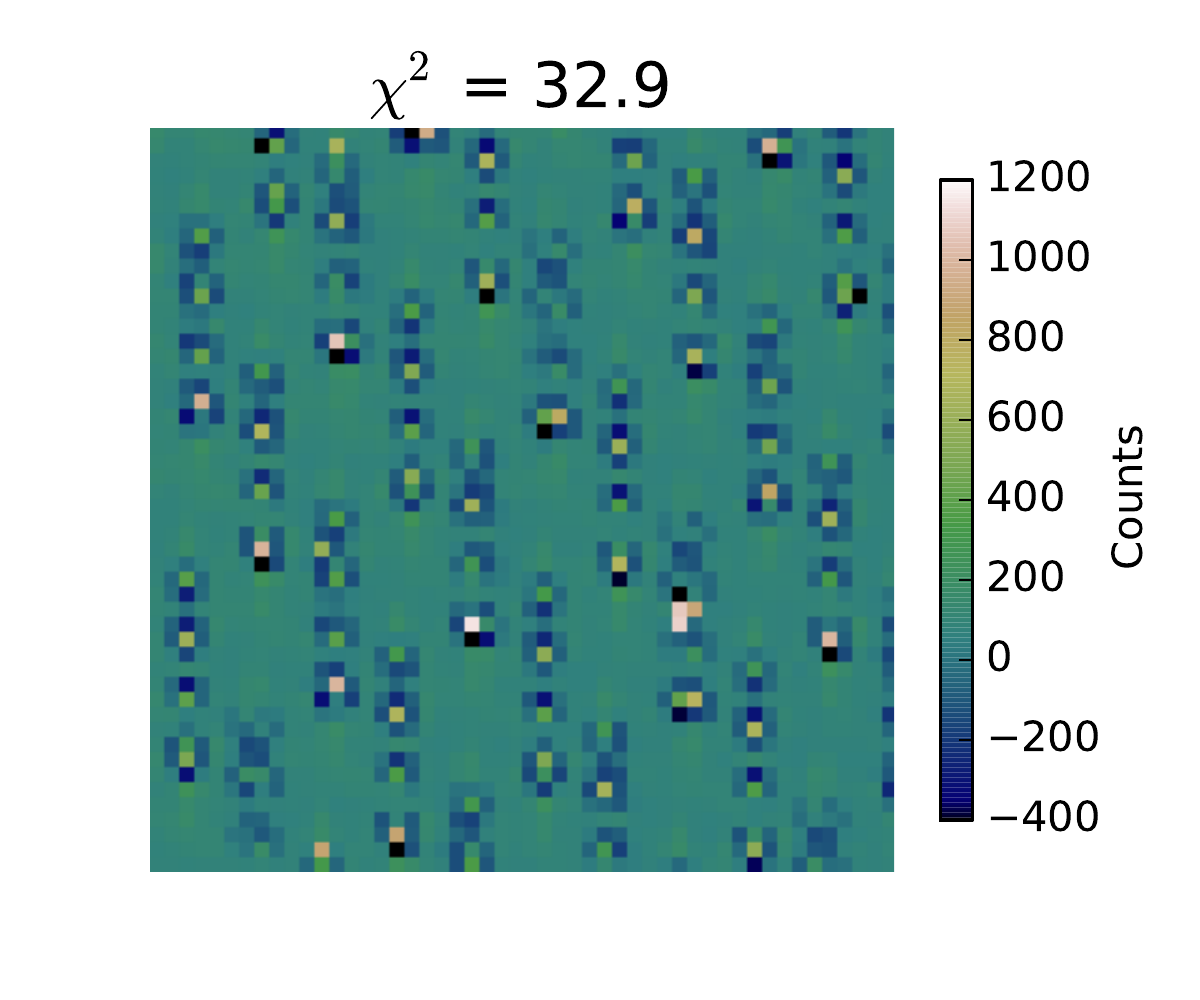}
   \label{fig:microlensres}
   }
   \subfigure[Microlens PSF Residuals]
    {
   \includegraphics[width=0.3\textwidth]{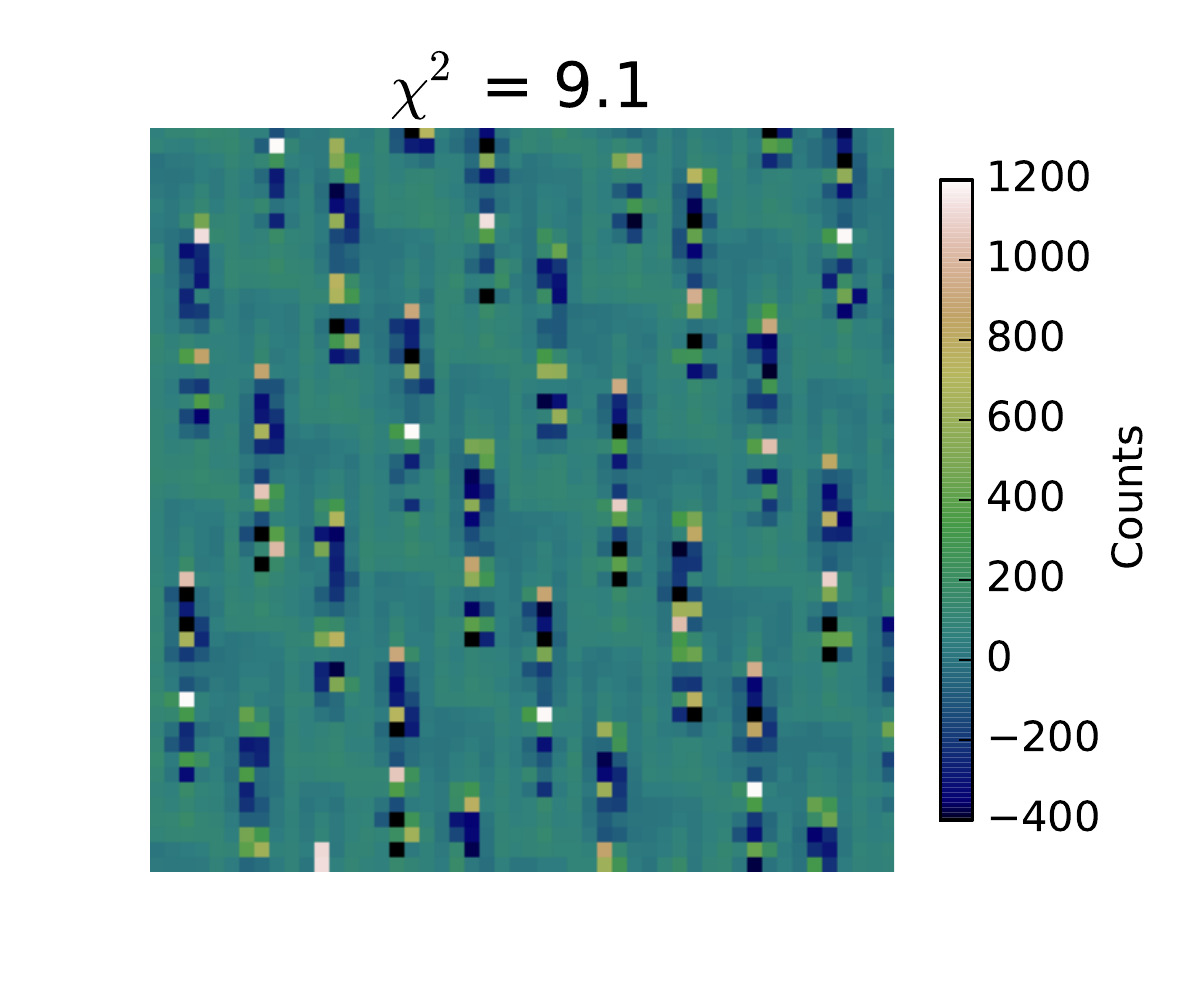}
   \label{fig:microlensres}
   }
   \caption[example] 
   { \label{fig:residuals} 
 (a) A 50 x 50 pixel cutout of an observed H band Xe arc lamp image. (b) A model Xe arc lamp image created using Gaussian PSFs. (c) A model Xe arc lamp image created using Microlens PSFs. (d) The residuals obtained by subtracting the observed lenslet spectrum array from a lenslet spectrum array created by simulating gaussian PSFs. The reduced $\chi^{2}$ value for the full image (2048 x 2048 pixels; ignoring bad pixels) is given. (e) The same as (d) for the microlens PSFs. Though both the Gaussian and microlens PSFs fit the peak locations, dispersion and spectral rotation well, the microlens method does a much better job of fitting the image background, and the shape of the PSF.}
   \end{figure}

There has been substantial work done to determine the shape of the microlens PSFs for this instrument by GPI team members (see Ingraham et. al., these proceedings).\cite{microlensspie} High resolution microlens PSFs for each lenslet have been produced in all bands. Use of these microlens PSFs in place of the Gaussian PSFs for the wavelength solution fitting was tested. Figure \ref{fig:residuals} shows residual spectral images using both the Gaussian and microlens PSFs to fit an \textit{H} band Xe arc lamp image. 
For the \textit{H} band Xe spectra, we find that the empirical microlens PSFs result in $\chi^{2}_{M} = 9.1$ (Reduced $\chi^{2}$ computed assuming the per-pixel noise $\sigma_{i}$ is given by the photon noise),  an improvement over the result using Gaussian PSFs of $\chi^{2}_{G} = 32.9$. However, when considering only the flux from the more brightly illuminated parts of the spectra by selecting the brightest third of the image pixels, the Gaussian PSFs provide a better fit with $\chi^{2}_{G} = 3.9$ and $\chi^{2}_{M} =  10.9$. 
This implies that the Gaussian PSFs provide a good fit to the cores of the emission line PSFs, though they do not fit as well the wings of the PSFs. The higher $\chi^{2}_{G}$ when computed over the full array is driven by the 68 \% of less-illuminated pixels between the spectra that are less used in the spectral extraction but are well fit by the wings of the microlens PSFs. The distribution of positions, dispersions, and tilts produced in both of these wavelength solutions are roughly the same. Because we are simultaneously fitting multiple spectral peaks at once, the pixel phase errors introduced in the Gaussian PSF fits average out and the mean position of each lenslet does not vary between the two methods. 
The main advantage of the microlens PSFs is an improvement in the accuracy of fitting a single peak and in distinguishing between blended peaks. For example, in $Y$ band, the systematic error in fitting a Gaussian PSF is $\sim$ 0.025 pixels, and for the microlens PSF it is $\sim$ 0.0004 pixels.\cite{microlensspie} This allows a more precise fit to the dispersion and tilts of the lenslets and will aid in future studies of the non-linearity of the wavelength solution. The microlens PSF implementation of the wavelength solution is not yet available in the public GPI Data Reduction Pipeline, but will be released for the 2014B observing semester.


\subsection{Quick Wavelength Algorithm}

The least squares fitting algorithm described above in Section \ref{sec:leastsqrs} works well, but is very computationally intensive and takes several hours to run. In order to calibrate small changes of the position of lenslet spectra on the detector at a fast timescale (i.e. for corrections due to flexure in the IFS as discussed in Section \ref{sec:flexure}), a quick wavelength solution algorithm was developed. The ``Quick Wavelength Solution" GPI Pipeline primitive calculates changes in position of the lenslet spectra from an arc lamp image by fixing all parameters except the $x_{0}$ and $y_{0}$ positions for a user-selectable subset of lenslets across the field of view and computing an average shift. By default, this primitive uses a grid of lenslets spaced twenty lenslets apart in row and column.  By combining information from multiple lenslets, we can achieve a good measurement of the $x$ and $y$ shifts with lower S/N arc lamp images to enable quick nighttime calibrations. This algorithm executes in seconds, several hundred times faster than the full wavelength calibration algorithm.

\section{WAVELENGTH PERFORMANCE}
\label{sec:results}

\subsection{Accuracy of the Wavelength Solution}

\begin{figure}
   \begin{center}
   \begin{tabular}{c}
   \includegraphics[height=5cm]{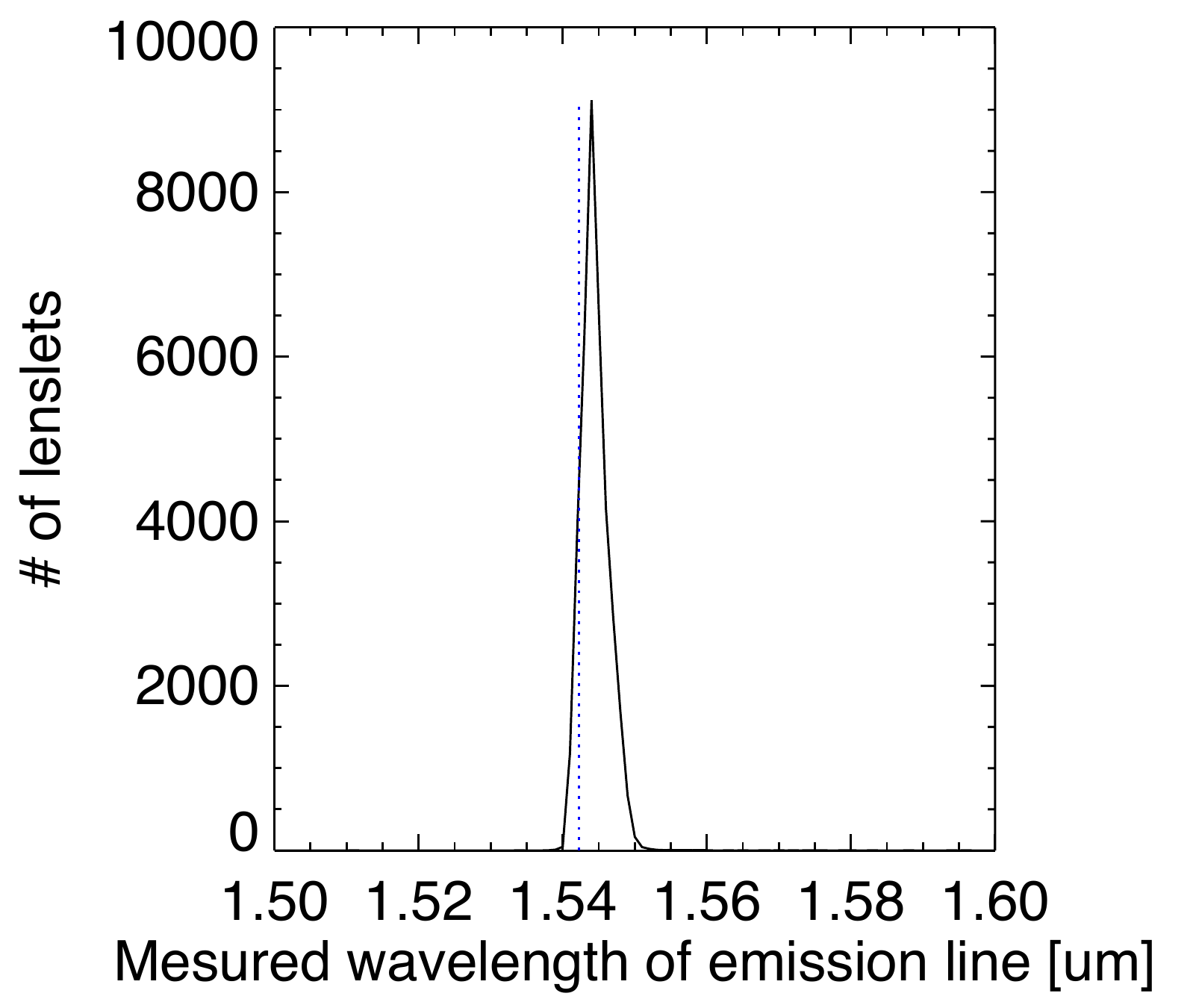}
   \end{tabular}
   \end{center}
   \caption[example] 
   { \label{fig:results} 
    Histogram of the measured wavelength in $\mu$m of the brightest spectral peak for an H band Xe arc lamp spectral extraction using an Ar wavelength solution, for all the lenslets in the image. The dotted line represents the theoretical location of the peak. The histogram is strongly peaked at a value within 0.032 pixels of the correct wavelength. }
   \end{figure}

To achieve the science goals of GPI, we require an uncertainty in the spectral characterization of $< 5\%$ which requires the wavelength solution to be accurate to within 1\%. To test the accuracy of the wavelength solution, we created an extracted datacube of a lamp image with 37 spectral channels and compared the theoretical location of the brightest spectral peak to a histogram of peak locations in the reduced cube. An example histogram using an \textit{H} band Ar wavelength solution to fit a Xe arc lamp is provided in Figure \ref{fig:results}. The histogram is sharply peaked at 0.0321 detector pixels from the expected location, demonstrating an accuracy in the wavelength solution of 0.032 \%. Note that the discrepancy of 0.0321 pixels is the value of pixel-phase error that you would expect from Gaussian fitting.\cite{microlensspie}  Table \ref{table} provides the accuracy in pixel location and percent (i.e. $\Delta \lambda / \lambda \times 100$) for all bands. In all bands, the peak wavelength was within a tenth of a pixel of the expected location and was accurate to within a tenth of a percent or less, well below the required accuracy. 

\begin{table}[h!]
\caption{Derived accuracy of the wavelength solution using the Ar arc lamp. Column 2 gives the difference in the measured and expected wavelength of the emission lines in detector pixel. Column 3 gives the associated wavelength discrepancy in microns. Column 4 gives the accuracy of the wavelength solution in percentage ($\Delta \lambda / \lambda \times 100$). } 
\label{table}
\begin{center}       
\begin{tabular}{|c|c|c|c|} 
\hline
\rule[-1ex]{0pt}{3.5ex}  Band & Pixel Offset & $\Delta \lambda \, (\mu m)$ & $\Delta \lambda / \lambda$ \% \\
\hline
\rule[-1ex]{0pt}{3.5ex}  Y & 0.096 & 0.0013 & 0.14   \\
\hline
\rule[-1ex]{0pt}{3.5ex}  J & 0.084 & 0.00068 & 0.054  \\
\hline
\rule[-1ex]{0pt}{3.5ex}  H & 0.032 & 0.00049 & 0.032  \\
\hline
\rule[-1ex]{0pt}{3.5ex}  K & 0.095 & 0.0014 & 0.07  \\
\hline
\end{tabular}
\end{center}
\end{table}

\subsection{Ar vs. Xe Lamps}
\label{sec:arvsxe}

\begin{figure}
   \begin{center}
   \includegraphics[width=1.0\textwidth]{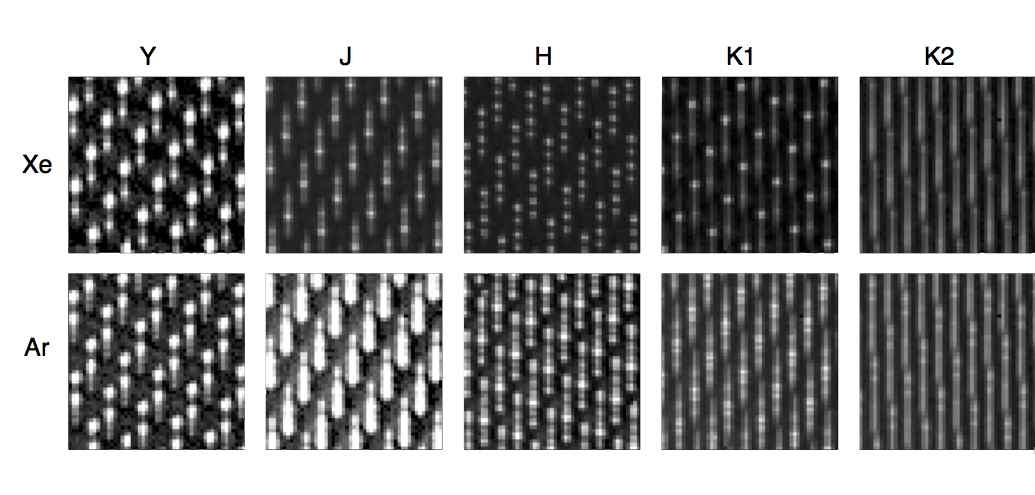}
   \end{center}
   \caption[example] 
   { \label{fig:alllamps} 
 Examples of GCAL spectra for Xe and Ar for all GPI filters, for a 50x50 pixel subregion near the center of the detector. These are all shown displayed on a log scale from 0.1 to 50 counts per second per detector pixel. The Ar spectra are consistently brighter than Xe but generally have less well separated emission lines. In the K1 and K2 spectra, the thermal background continuum is visible and at K2 is the dominant source of light. Background exposures must be observed and subtracted prior to generating wavecals for K1 and K2 but are not needed at shorter wavelengths.  }
   \end{figure}

The accuracy of the wavelength solution is limited by the calibration sources available in GCAL. At the spectral resolution of GPI, neither the Xe nor the Ar lamps provide multiple defined peaks to fit the spectral dispersion in all bands.  Figure \ref{fig:alllamps} provides a cutout of the detector images for both Xe and Ar arc lamps in all bands.  To fit the wavelength solution as described in Eq. (1), a spectrum must have multiple sharp and unblended peaks. Multiple blended peaks at low signal to noise bias the dispersion estimate and consequently, the spectral positions. For all bands but Y, the Xe lamp provided the most accurate solution. However, the Xe lamp in GCAL is $\sim$ 3 times fainter than the Ar lamp, requiring more time spent on overhead calibrations. With the new least squares algorithm, we are able to reproduce the results of the Xe lamp with the brighter Ar lamp by fitting many of the blended lines at once. 
 The GPI pipeline is able to produce wavelength solutions with $< 1\%$ uncertainty for both lamps.

\begin{figure}
   \begin{center}
   \includegraphics[width=1.0\textwidth]{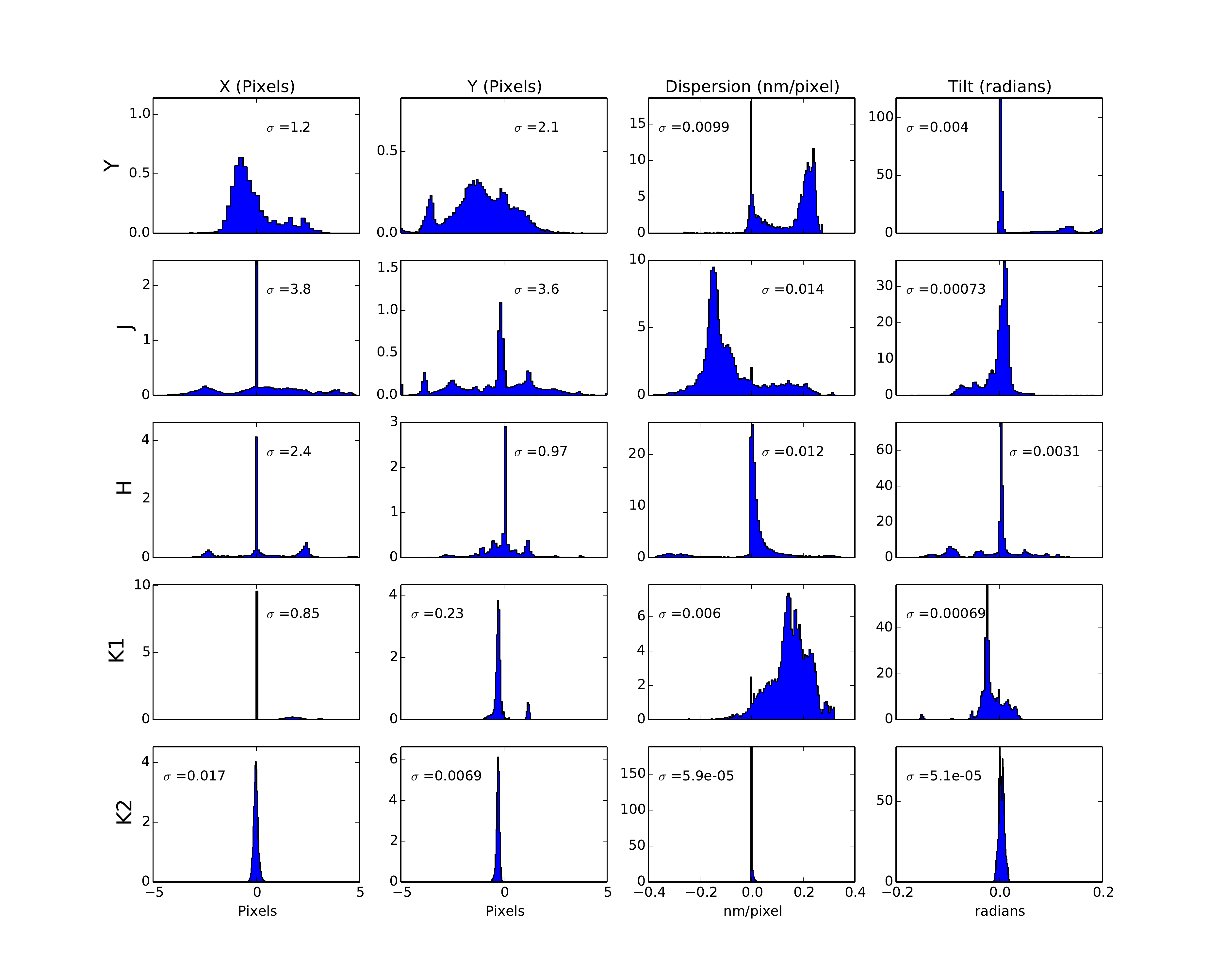}
   \end{center}
   \caption[example] 
   { \label{fig:arxe} 
   Performance of the Ar and Xe lamps. This figure gives histograms of the difference in the $x$ and $y$ positions, dispersions ($w$) and tilt ($\theta$) values between the Ar and Xe lamps for each band. The bands are listed down the side and the spectral properties are listed across the top. Histograms were calculated using Xe - Ar values.  The probability distribution (sum of the bin size times the number of lenslets in that bin) is normalized to one, with 100 bins per histogram. The variance $\sigma$ is included with each plot. Differences between the Ar and Xe lamp wavelength calibrations are generally small, but there are biases in the solutions in the \textit{Y, J,} and \textit{K1} bands caused by the different sets of emission lines available between the two lamps.   }
   \end{figure}

Figure \ref{fig:arxe} examines the disparities in the wavelength solution produced by the Ar and Xe lamps. Wavelength solutions for each band were calculated separately using both the Ar and Xe arc lamps. Histograms of the Xe - Ar values for the $x$ and $y$ positions, dispersions ($w$) and spectral tilt ($\theta$) are given. The Ar and Xe wavelength solutions agree best in \textit{K2} band with $\sigma <  0.02$, where both arc lamps have only two distinct peaks. The consequences of mixed emission lines is demonstrated well in the \textit{J} band dispersion histogram in Figure \ref{fig:arxe}. \textit{J} band has only one easily distinguishable line and a faint clump of blended lines in both Xe and Ar. Due to uncertainties in the best fit location of the fainter, blended lines, the dispersion preferred by the Xe lamp wavelength solution is $\sim 0.2$ nm/pixel greater than the Ar lamp solution, leading to a disagreement in the $x_{0}$ and $y_{0}$ locations. The x-axes of the histograms exacerbate the disagreements between the Xe and Ar lamps, however, the $\sim 0.2$ nm/pixel discrepancy in dispersion only contributes to a $\sim$ 1 \% uncertainty (average dispersion is $\sim$ 14 nm/pixel).

\section{FLEXURE}
\label{sec:flexure} 
	
\begin{figure}
   \begin{center}
   \begin{tabular}{c}
   \includegraphics[height=18cm]{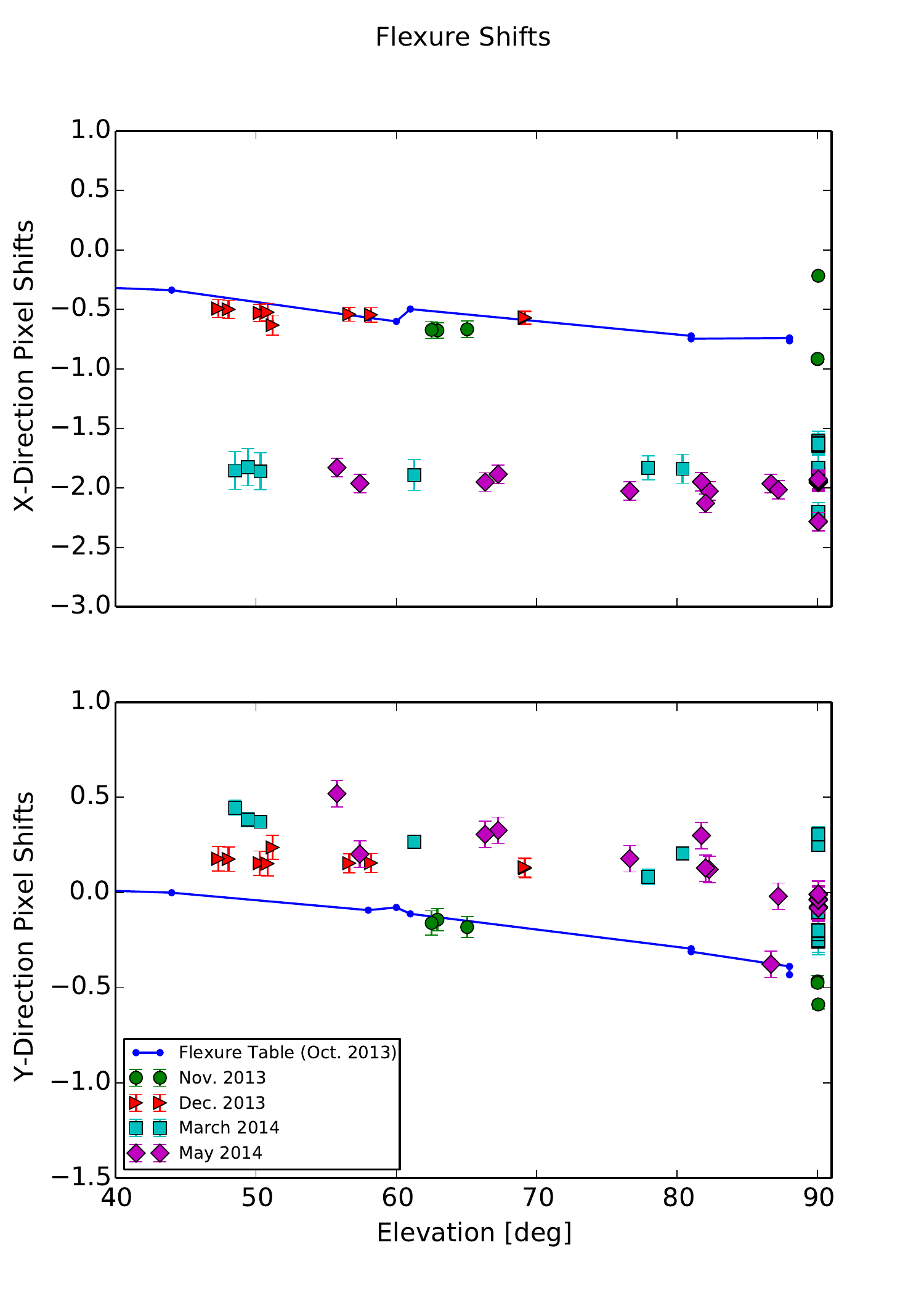}
   \end{tabular}
   \end{center}
   \caption[example] 
   { \label{fig:flexure} 
   Gravitationally induced flexure shifts with varying elevation for the four GPI commissioning runs thus far. Top: x-shifts (roughly perpendicular to dispersion direction) as a function of elevation.  Bottom: y-shifts (parallel to dispersion direction) as a function of elevation. The predicted shifts from the Flexure table calibration file constructed from October 2013 data is given by the blue dashed line. Data taken on different dates is color coded and described by the legend. During the course of a single night, the x and y shifts with elevation are repeatable. However, large shifts occur from night to night most likely due to rotation of GPI about it's rotational axis while other Gemini Instruments are in use. }
   \end{figure}

The wavelength calibration algorithm described above is capable of tracking the motion of the lenslet spectra across the detector plane to an accuracy of 1/10th of a pixel. Therefore, it can be used to trace the flexure of the optics within the IFS causing shifts in the spectral positions with instrument position. These shifts are thought to be caused by motion of one or more of the optics between the lenslet array and the detector. Using the flexure rig at Gemini South Observatory during early commissioning, we examined the magnitude and direction of shifts due to IFS flexure from motions of GPI when the telescope moves in elevation, and from rotations of the Cassegrain Rotator about the telescope optical axis. 
The observed flexure appears to be a complex function of instrument current elevation, hysteresis from prior elevations, and occasional larger shifts which sometimes but not always correlate with thermal cycling of the IFS. ÊThere is a reproducible general trend as a function of elevation but substantial scatter around this due to changing offsets as a function of time. These factors are not yet all fully understood.
With changes in elevation from zenith to the horizon, the observed flexure follows an arc showing $\sim$0.8 pixels of motion along the X-axis and $\sim$0.4 pixels along the Y-axis. The motion is generally repeatable to within 0.1 pixels.
Figure \ref{fig:flexure} shows the change in position on the detector over time due to flexure. The bulk shifts between observing runs are thought to be partially the result of motion about the rotational axis which occurs when other Gemini instruments that require compensation for field rotation are in use. Over the course of an observing run, only the elevation axis is expected to change with elevation of the target. The rotational axis is only affected when other instruments on Gemini South are observing. 

\begin{figure}
   \begin{center}
   \begin{tabular}{c}
   \includegraphics[height=7cm]{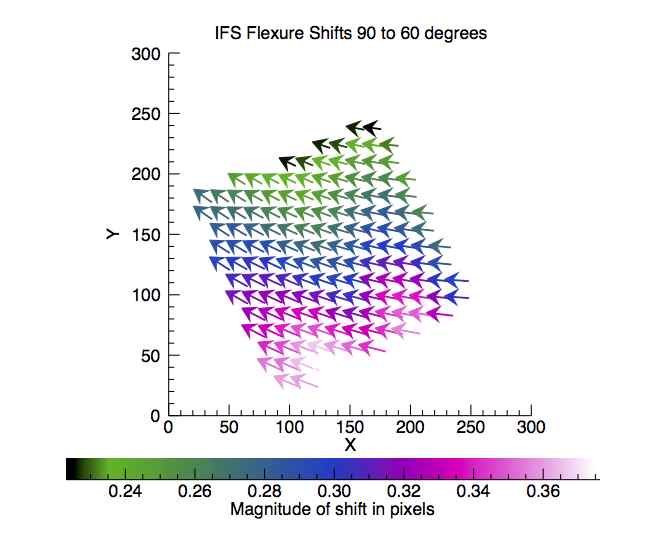}
   \end{tabular}
   \end{center}
   \caption[example] 
   { \label{fig:fovflexure} 
Flexure across the field of view. Using arc lamp images taken at zenith (90 degrees) and 30 degrees off of zenith, we track the variation in flexure over the detector area. There is a clear variation from top to bottom of the FOV, but at 30 degrees, the difference is $\pm$ 0.08 pixels. }
   \end{figure}

The shifts due to flexure are not constant across the field of view of the detector. Figure \ref{fig:fovflexure} provides a vector plot illustrating the change in position of the lenslets over the detector resulting from a 30 degree change in elevation. The magnitude of the flexure shifts changes by $\sim$ 0.15 pixels across the detector. At most, this will cause flexure variations of $\sim 0.08$ pixels from the mean shifts in $x$ and $y$ for an image which is below the threshold for uncertainties in the wavelength solution. Thus, deviations in the flexure correction for different regions of the detector are currently being ignored. It is sufficient to use only the mean shifts in the correction. In the future it may be possible to correct for flexure within the spectral extraction process by implementing a least square inversion flux extraction method e.g. Maire et. al. and Draper et. al. (this proceedings).\cite{fluxextraction,jerome}

\section{Recommended Practices}

Deep arc lamp exposures should be taken in all bands at least once an observing semester to calibrate the GPI Integral Field Spectrograph with a spectral accuracy of $< 1$\%. This requires Xe or Ar arc lamp images with SNR $\gg 20$ per pixel in the emission line wings, corresponding to SNR of $\sim$ 50 - 80 at the spectral peaks. In \textit{H} and both \textit{K} bands an Ar arc lamp is sufficient and requires less time spent on calibrations. We recommend using the Xe arc lamp for \textit{J} and \textit{Y} bands because the Ar lamp does not have sufficient bright and unblended peaks to perform an accurate calibration. To correct the spectral positions for flexure variations with elevation and from night to night, a single one-minute \textit{H} band Ar arc lamp exposure is recommended contemporary with each science target, at the same elevation. 

The GPI Data Reduction Pipeline (See Perrin et. al., these proceedings)\cite{perrinpipe} includes modules (termed primitives) to create both the high S/N wavelength calibration files and a fast method for determining offsets from a short arc lamp exposure. The ``2D Wavelength Solution" primitive performs the full wavelength solution for all lenslets. Because that this primitive is computationally intensive and takes several hours to run, the ``Quick Wavelength Solution" primitive was developed to fit only the positions of a subset of the lenslet spectra over the field of view to calculate an average bulk shift. For the quick look reductions produced at Gemini, If an arc lamp image in any band is taken directly before a science image, the GPI DRP will automatically run the quick wavelength solution algorithm, determine the positional shifts due to flexure, extrapolate those shifts to the band of the science observations and correct for these shifts when performing the spectral extraction. If an arc lamp image isn't taken prior to a science image, the pipeline will use the most recent arc lamp image for the correction.

\acknowledgments     
 
This material is based upon work supported by the National Science Foundation Graduate Research Fellowship under Grant No. DGE-1232825. Any opinion, findings, and conclusions or recommendations expressed in this material are those of the authors and do not necessarily reflect the views of the National Science Foundation. The Gemini Observatory is operated by the Association of Universities for Research in 
Astronomy, Inc., under a cooperative agreement with the NSF on behalf of the Gemini 
partnership: the National Science Foundation (United States), the National Research 
Council (Canada), CONICYT (Chile), the Australian Research Council (Australia), 
Minist\'erio da Ci\'encia, Tecnologia e Inova\c{c}\=ao (Brazil), and Ministerio de Ciencia, 
Tecnolog\'ia e Innovaci\'on Productiva (Argentina). Portions of this work were supported by the STScI Director's Discretionary Research Fund. 

\bibliographystyle{spiebib}   
\bibliography{report}   

\begin{thebibliography}{10}

\bibitem{larkin}
{James E. Larkin}, {Jeffrey K. Chilcote}, {Theodore Aliado}, {Brian J. Bauman},
  {George Brims}, {John M. Canfield}, {Andrew Cardwell}, {Daren Dillon}, {Rene
  Doyon}, {Jennifer Dunn}, {Michael P. Fitzgerald}, {James R. Graham}, {Stephen
  Goodsell}, {Markus Hartung}, {Pascale Hibon}, {Patrick Ingraham},
  {Christopher A Johnson}, {Evan Kress}, {Quinn M. Konopacky}, {Bruce A.
  Macintosh}, {Kenneth G. Magnone}, {Jerome Maire}, {Ian S. McLean}, {David
  Palmer}, {Marshall D. Perrin}, {Carlos Quiroz}, {Fredrik Rantakyrš}, {Naru
  Sadakuni}, {Leslie Saddlemyer}, {Andrew Serio}, {Simon Thibault}, {Sandrine
  J. Thomas}, {Philippe Vallee}, and {Jason L. Weiss}, ``{The Integral Field
  Spectrograph for the Gemini Planet Imager},'' {\em Society of Photo-Optical
  Instrumentation Engineers (SPIE) Conference Series} {\bf 9147} (2014).

\bibitem{2014AAS22320204C}
{Chilcote}, J.~K., {Larkin}, J.~E., {Planet Imager instrument}, G., and
  {science Teams}, ``{Development and Commissioning of the Integral Field
  Spectrograph for the Gemini Planet Imager},'' in [{\em American Astronomical
  Society Meeting Abstracts}{\nolinebreak\hspace{0.1em}]},  {\em American
  Astronomical Society Meeting Abstracts} {\bf 223},  202.04 (Jan. 2014).

\bibitem{2011PASP123692M}
{McBride}, J., {Graham}, J.~R., {Macintosh}, B., {Beckwith}, S.~V.~W.,
  {Marois}, C., {Poyneer}, L.~A., and {Wiktorowicz}, S.~J., ``{Experimental
  Design for the Gemini Planet Imager},'' {\em PASP}~{\bf 123},  692--708 (June
  2011).

\bibitem{2014AAS22334814P}
{Perrin}, M.~D., {Gemini Planet Imager instrument}, T., and {science Teams},
  ``{Gemini Planet Imager Data Analysis Methods, Software, and First Data
  Release},'' in [{\em American Astronomical Society Meeting
  Abstracts}{\nolinebreak\hspace{0.1em}]},  {\em American Astronomical Society
  Meeting Abstracts} {\bf 223},  348.14 (Jan. 2014).

\bibitem{perrinpipe}
Perrin, M., Maire, J., Ingraham, P.~J., Savransky, D., Millar-Blanchaer, M.,
  Wolff, S.~G., Ruffio, J.-B., Wang, J.~J., Draper, Z.~H., Sadakuni, N.,
  Marois, C., Rajan, A., Fitzgerald, M.~P., Macintosh, B., Graham, J.~R.,
  Doyon, R., Larkin, J.~E., Chilcote, J.~K., Goodsell, S.~J., Palmer, D.~W.,
  Labrie, K., Beaulieau, M., Rosa, R. J.~D., Greenbaum, A.~Z., Hartung, M.,
  Hibon, P., Konopacky, Q.~M., Lafreniere, D., Lavigne, J.-F., Marchis, F.,
  Patience, J., Pueyo, L.~A., Rantakyro, F., Soummer, R., Sivaramakrishnan, A.,
  Thomas, S.~J., Ward-Duong, K., and Wiktorowicz, S., ``Gemini planet imager
  observational calibrations i: overview of the gpi data reduction pipeline,''
  {\em Society of Photo-Optical Instrumentation Engineers (SPIE) Conference
  Series} {\bf 9147} (2014).

\bibitem{1997SPIE.2871.1171R}
{Ramsay Howat}, S.~K., {Harris}, J.~W., and {Bennett}, R.~J., ``{Dedicated
  calibration facility for the Gemini Telescopes},'' in [{\em Optical
  Telescopes of Today and Tomorrow}{\nolinebreak\hspace{0.1em}]},  {Ardeberg},
  A.~L., ed., {\em Society of Photo-Optical Instrumentation Engineers (SPIE)
  Conference Series} {\bf 2871},  1171--1178 (Mar. 1997).

\bibitem{2014AAS22322902M}
{Macintosh}, B., {Gemini Planet Imager instrument Team}, {Planet Imager
  Exoplanet Survey}, G., and {Observatory}, G., ``{The Gemini Planet Imager},''
  in [{\em American Astronomical Society Meeting
  Abstracts}{\nolinebreak\hspace{0.1em}]},  {\em American Astronomical Society
  Meeting Abstracts} {\bf 223},  229.02 (Jan. 2014).

\bibitem{2014arXiv1403.7520M}
{Macintosh}, B., {Graham}, J.~R., {Ingraham}, P., {Konopacky}, Q., {Marois},
  C., {Perrin}, M., {Poyneer}, L., {Bauman}, B., {Barman}, T., {Burrows}, A.,
  {Cardwell}, A., {Chilcote}, J., {De Rosa}, R.~J., {Dillon}, D., {Doyon}, R.,
  {Dunn}, J., {Erikson}, D., {Fitzgerald}, M., {Gavel}, D., {Goodsell}, S.,
  {Hartung}, M., {Hibon}, P., {Kalas}, P.~G., {Larkin}, J., {Maire}, J.,
  {Marchis}, F., {Marley}, M., {McBride}, J., {Millar-Blanchaer}, M.,
  {Morzinski}, K., {Norton}, A., {Oppenheimer}, B.~R., {Palmer}, D.,
  {Patience}, J., {Pueyo}, L., {Rantakyro}, F., {Sadakuni}, N., {Saddlemyer},
  L., {Savransky}, D., {Serio}, A., {Soummer}, R., {Sivaramakrishnan}, A.,
  {Song}, I., {Thomas}, S., {Wallace}, J.~K., {Wiktorowicz}, S., and {Wolff},
  S., ``{The Gemini Planet Imager: First Light},'' {\em ArXiv e-prints}  (Mar.
  2014).

\bibitem{testgpipaper}
{Maire}, J., {Perrin}, M.~D., {Doyon}, R., {Chilcote}, J., {Larkin}, J.~E.,
  {Weiss}, J.~L., {Marois}, C., {Konopacky}, Q.~M., {Millar-Blanchaer}, M.,
  {Graham}, J.~R., {Dunn}, J., {Galicher}, R., {Marchis}, F., {Wiktorowicz},
  S.~J., {Labrie}, K., {Thomas}, S.~J., {Goodsell}, S.~J., {Rantakyro}, F.~T.,
  {Palmer}, D.~W., and {Macintosh}, B.~A., ``{Test results for the Gemini
  Planet Imager data reduction pipeline},'' in [{\em Society of Photo-Optical
  Instrumentation Engineers (SPIE) Conference
  Series}{\nolinebreak\hspace{0.1em}]},  {\em Society of Photo-Optical
  Instrumentation Engineers (SPIE) Conference Series} {\bf 8451} (Sept. 2012).

\bibitem{Levenberg}
{Levenberg}, K., ``{A Method for the Solution of Certain Non-Linear Problems in
  Least Squares},'' {\em The Quarterly of Applied Mathematics}~{\bf 2},
  164--168 (1963).

\bibitem{marquardt}
{Marquardt}, D., ``{An algorithm for least-squares estimation of nonlinear
  parameters},'' {\em Journal of the Society for Industrial and Applied
  Mathematics}~{\bf 11},  431--441 (1944).

\bibitem{microlensspie}
Ingraham, P.~J., Ruffio, J.-B., Perrin, M.~D., Wolff, S., Draper, Z., Maire,
  J., Marchis, F., and Fesquet, V., ``Gemini planet imager observational
  calibrations iii: Empirical measurement methods and applications of
  high-resolution microlens psfs,'' {\em Society of Photo-Optical
  Instrumentation Engineers (SPIE) Conference Series} {\bf 9147} (2014).

\bibitem{fluxextraction}
{Draper, Zachary H.} and the GPI~Team, ``{Gemini Planet Imager Observational
  Calibrations IX: Least Square Inversion Flux Extraction},'' {\em SPIE
  \textbf{these proceedings}} (2014).

\bibitem{jerome}
{Maire, J\'{e}r\^{o}me} and the GPI~Team, ``{Gemini Planet Imager Observational
  Calibrations VI: Photometric and Spectroscopic Calibration for the Integral
  Field Spectrograph},'' {\em SPIE \textbf{these proceedings}} (2014).

\end{thebibliography}

\end{document}